\documentclass[twocolumn,preprintnumbers,nofootinbib]{revtex4-1}
\usepackage{graphicx}
\usepackage{amssymb}
\usepackage{color}
\def\beq{\begin{equation}}
\def\eeq{\end{equation}}
\def\beqn{\begin{eqnarray}}
\def\eeqn{\end{eqnarray}}

\begin{document}
\title{A model for non-singular black hole collapse and evaporation}
\author{Sabine Hossenfelder${}^1$, Leonardo Modesto$^{2}$, 
Isabeau Pr\'emont-Schwarz${}^2$}
\affiliation{${}^1$Nordita, Roslagstullsbacken 23, 106 91 Stockholm, Sweden \\
${}^2$Perimeter Institute for Theoretical Physics, 31 Caroline St.N., Waterloo, ON N2L 2Y5, Canada 
}

\begin{abstract}
We study the formation of a black hole and its subsequent evaporation in
a model employing a minisuperspace approach to loop quantum gravity. In
previous work the static solution was obtained and shown to
be singularity-free. Here, we examine the more realistic dynamical
case by generalizing the static case with help of the Vaidya metric.
We track the formation and evolution of trapped surfaces during collapse
and evaporation and examine the buildup of quantum gravitationally
caused stress-energy preventing the formation of a singularity.
\end{abstract}
\maketitle

\section{Introduction}

 The formation of black hole singularities is an
inevitable consequence of General Relativity. As instances of infinite energy
density and tidal forces, black holes have made headlines, inspired science fiction
movies, and were studied in thousands of research articles. It adds to
the fascination that we know today black holes are not just a mathematically
possible solution to Einstein's field equations, but part of Nature.
Since more than a decade now, we have good evidence that our Milky Way, as other
galaxies, hosts many stellar black holes as well as a supermassive black hole in its
center.

From the perspective of quantum gravity, black holes are of interest because of
the infinite curvature towards their center which signals a
breakdown of General Relativity. It is an area where effects of quantum
gravity are strong, and it is generally expected that these effects
prevent the formation of the singularity. Since the black hole emits
particles in the process of Hawking radiation \cite{Haw1}, the horizon
radius decreases. In the standard case it approaches the singularity
until both, the singularity and the horizon, vanish in the
endpoint of evaporation \cite{Haw2}. However, if the singularity does not
exist, this scenario cannot be correct. Since the singularity plays
a central role for the causal space-time diagram, its absence in the
presence of quantum gravitational effects has consequences for the
entire global structure \cite{AB}, and the removal of the singularity is
essential for resolving the black hole information loss problem \cite{Infloss}. To understand the dynamics of
the gravitational and matter fields, it is then necessary to have a concrete model.

It is thus promising that it has been shown in a simplified version
of loop quantum gravity, known as loop quantum cosmology (LQC) \cite{Bojowald:2006da}, a
resolution of singularities, the big bang as well as the black
hole singularity \cite{Ashtekar:2005qt,M1,M2}, can be achieved. The regular static
black hole metric was recently derived in \cite{Modesto:2008im}, and
studied more closely in \cite{Modesto:2009ve}.  A resolution of the black hole 
singularity was also obtained in an effective, noncommutative approach to quantum gravity 
\cite{Nicolini:2005vd} and in asymptotically safe quantum gravity \cite{RB}.
In another work \cite{Ashtekar:2008jd}, a
2-dimensional model was used to study the evaporation process in the absence
of a singularity. 

Here, we will use a 4-dimensional model based on the static solution
derived in \cite{Modesto:2008im} and generalize it to a dynamical case
which then allows us to examine the causal structure. This generalization
holds to good accuracy in all realistic scenarios. This approach
should be understood not as an exact solution to a problem that requires
knowledge of a full theory of quantum gravity, but as a plausible model based on preliminary studies that allows us to investigate the
general features of such regular black hole solutions.

Non-singular black holes were considered already by
Bardeen in the late 60s and have a long history \cite{Bar,Frolov:1989pf,Frolov:1988vj,Balbinot:1990zz,Aurilia:1990uq,Bar2,Bar3,Bar4,Bar5,Bar6,Hayward:2005gi,Nicolini:2005zi,Ansoldi:2006vg,Spallucci:2008ez,Nicolini:2009gw,Nicolini:2008aj}. We will here
use a procedure similar to that in \cite{Hayward:2005gi}. The rest
of the paper is organized as follows. We start in the next section
by recalling the regular static metric we will be using. In section
\ref{dynamical} we generalize it to a collapse scenario and discuss
its properties. In section \ref{thermo} we summarize the thermodynamical
properties and, in section \ref{evap}, add the evaporation
process and construct the complete causal diagram.
The signature of the metric is $(-,+,+,+)$ and we use the unit
convention $\hbar=c=G_N=1$.

\section{The Regular Schwarzschild-metric}
\label{static}

Let us first summarize the regular black hole metric that we will be
using.

Loop Quantum Gravity (LQG) is a candidate theory of quantum gravity. It is obtained from the canonical quantization of the Einstein equations written in terms of the Ashtekar variables \cite{AA}, that is in terms of an $\mathrm{su}(2)$ 3-dimensional connection $A$ and a triad $E$. The result \cite{Mercuri:2010xz} is that the basis states of LQG are closed graphs the edges of which are labelled by irreducible $\mathrm{su}(2)$ representations and the vertices by $\mathrm{su}(2)$ intertwiners. Physically, the edges represent quanta of area with area $\gamma l_{\rm P}^2 \sqrt{j(j+1)}$, where $j$ is the representation label on the edge (a half-integer), $l_{\rm P}$ is the Planck length,  and $\gamma$ is a parameter of order 1 called the Immerzi parameter. Vertices of the graph represent quanta of 3-volume. The important observation to make here is that area is quantized and the smallest quanta of area possible has area $\sqrt{3}/2 \gamma l_{\rm P}^2$. 

The regular black hole metric that we will be using is derived from a simplified model of LQG \cite{Modesto:2008im}. To obtain this simplified model we make the following assumptions. First of all, the number of variables is reduced by assuming spherical symmetry. Then, instead of all possible closed graphs, a regular lattice with edge lengths $\delta_1$ and $\delta_2$ is used. The solution is then obtained dynamically inside the homogeneous region (inside the horizon where space is homogeneous but not static). Analytically continuing the solution outside the horizon one finds that one can reduce the two free parameters by imposing that the minimum area present in the solution corresponds to the minimum area of LQG. The one remaining unknown constant $\delta$ is a parameter of the model determining the strength of deviations from the classical theory, and would have to be constrained by experiment. With the plausible expectation that the quantum graviational corrections become relevant only when the curvature is in the Planckian regime, corresponding to $\delta < 1$, outside the horizon the solution is the Schwarzschild solution up to negligible Planck-scale corrections which allows us to believe the legitimacy of the analytical extension outside the horizon. 

This quantum gravitationally corrected
Schwarzschild metric can be expressed in the form
\begin{eqnarray}
ds^2 = - G(r) dt^2 + \frac{dr^2}{F(r)} + H(r) d\Omega~,
\label{g}
\end{eqnarray}
with $d \Omega = d \theta^2 + \sin^2 \theta d \phi^2$ and
\begin{eqnarray}
&& G(r) = \frac{(r-r_+)(r-r_-)(r+ r_{*})^2}{r^4 +a_0^2}~ , \nonumber \\
&& F(r) = \frac{(r-r_+)(r-r_-) r^4}{(r+ r_{*})^2 (r^4 +a_0^2)} ~, \nonumber \\
&& H(r) = r^2 + \frac{a_0^2}{r^2}~ .
\label{statgmunu}
\end{eqnarray}
Here, $r_+ = 2m$ and $r_-= 2 m P^2$ are the two horizons, and $r_* = \sqrt{r_+ r_-} = 2mP$. $P$ is the
polymeric function $P = (\sqrt{1+\epsilon^2} -1)/(\sqrt{1+\epsilon^2} +1)$, with
$\epsilon \ll 1$ the product of the Immirzi parameter ($\gamma$) and the polymeric parameter ($\delta$). With this, it is 
also $P \ll 1$, such that $r_-$ and $r_*$ are very close to $r=0$. The area $a_0$ is equal to $A_{\rm min}/8 \pi$, $A_{\rm min}$ being the minimum area gap of LQG.

Note that in the above metric, $r$ is only asymptotically the usual radial
coordinate since $g_{\theta \theta}$ is not just $r^2$. This choice of
coordinates however has the advantage of easily revealing the properties
of this metric as we will see. But first, most importantly, in the limit
$r \to \infty$ the deviations from the Schwarzschild-solution are of
order $M \epsilon^2/r$, where $M$ is the usual ADM-mass:
\beqn
G(r) &\to& 1-\frac{2 M}{r} (1 - \epsilon^2)~, \nonumber  \\
F(r) &\to& 1-\frac{2 M}{r}~ , \nonumber \\
H(r) &\to& r^2 .
\eeqn
The ADM mass is the mass inferred by an observer at flat asymptotic infinity; it is determined solely 
by the metric at asymptotic infinity.  The parameter $m$ in the solution is related to the mass $M$ by $M = m (1+P)^2$.

If one now makes the coordinate transformation $R = a_0/r$ with the rescaling 
$\tilde t= t \, r_*^{2}/a_0$, and
simultaneously substitutes $R_\pm = a_0/r_\mp$, $R_* = a_0/r_*$ one finds that the metric in
the new coordinates has the same form as in the old coordinates and thus exhibits a
very compelling type of self-duality with dual radius $r=\sqrt{a_0}$. Looking at the angular part
of the metric, one sees that this dual radius corresponds to a minimal possible 
surface element. It is then also clear that in the limit $r\to 0$, corresponding
to $R\to \infty$, the solution
does not have a singularity, but instead has another asymptotically flat Schwarzschild region.

\begin{figure}
\includegraphics[width=8cm]{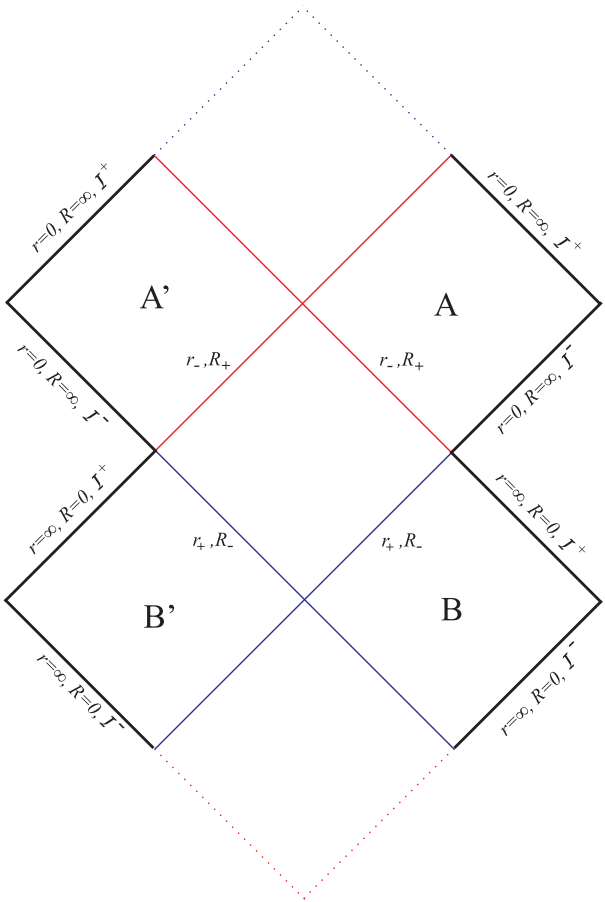}
\caption{Penrose diagram of the regular static black hole solution with two
asymptotically flat regions. The both horizons, located at $r_+$ and $r_-$,
are marked in blue and red respectively. }
\label{bhh0}
\end{figure}

The causal diagram for this metric, shown in Fig \ref{bhh0}, then has two horizons and two pairs of asymptotically
flat regions, $A, A'$ and $B,B'$, as opposed to one such pair in the standard case. In the region enclosed by 
the horizons, space- and timelikeness
is interchanged. The horizon at $r_+$ is a future horizon for observers in the asymptotically flat $B,B'$ region  and a past horizon for observers inside the two horizons. Similarly, the $r_-$ horizon is a future horizon for observers inside the two horizons but a past horizon for observes in $A, A'$.  If one computes the time
it takes for a particle to reach $r=0$, one finds that it takes infinitely long \cite{Modesto:2009ve}. The diagram shown in Fig \ref{bhh0} is not analytically complete, but should be read as being
continued on the dotted horizons at the bottom and top. 


The metric in Eq. (\ref{statgmunu}) is a solution of a quantum gravitationally corrected set of
equations which, in the absence of quantum corrections $\epsilon, a_0 \to 0$, reproduce Einstein's field equations.
However, due to these quantum corrections, the
above metric is no longer a vacuum-solution to Einstein's field
equations. Instead, if one computes the Einstein-tensor and sets it
equal to a source term $G_{\mu \nu} = 8 \pi \widetilde{T}_{\mu \nu}$, one
obtains an effective quantum gravitational stress-energy-tensor $\widetilde{T}_{\mu \nu}$.
The exact expressions for the components of $\widetilde{T}$ are
somewhat unsightly and can
be found in the appendix. 
For our purposes it is here
sufficient to note that the entries are not positive definite and
violate the positive energy condition which is one of the assumptions
for the singularity theorems.

\section{Collapse}
\label{dynamical}

We will proceed by combining the static metric with a radially ingoing null-dust,
such that we obtain a dynamical space-time for a black hole formed
from such dust. Usually described by the Vaidya metric \cite{Vai}, we will
in this scenario have corrections to the Vaidya metric that are negligible in the
asymptotic region, but avoid the formation of a singularity in the strong-curvature
region. The metric constructed this way in the following is not a strict solution of
the minusuperspace LQC equations. However, as long as the null-dust does not
already display strong quantum gravitational effects by its mass profile,
this solution should hold to good accuracy\footnote{It has been claimed in \cite{Barcelo:2007yk}
that, counterintuitively, quantum gravitational effects could become important already
at the horizon when the collapse proceeds slowly. However, since we are considering
null-dust, the collapse is as fast as can possibly be and these considerations do not
apply.}.

\begin{figure}
\hspace{-0.8cm}
\includegraphics[width=8cm]{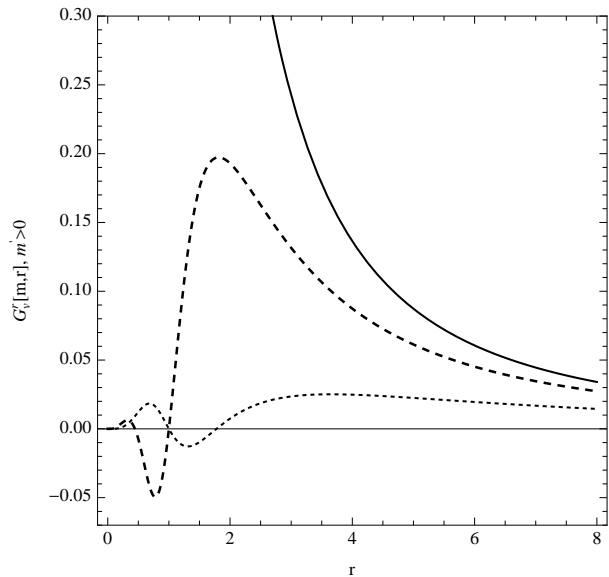}
\caption{$G^r_{\;v}$ as a function of $r$ for radially ingoing radiation and $m'(v) = 1$. The solid line depicts the classical case
for $\epsilon, a_0 \to 0$. The
long dashed line is for $m(v) = 20$, ($r_* > \sqrt{a_0}$) and the short dashed line is for $m(v)=5$, ($r_* < \sqrt{a_0}$). All quantities are in Planck units.}
\label{radiation}
\end{figure}

We start by making a coordinate transformation and rewrite the static space-time in terms
of the ingoing null-coordinate $v$. It is defined by the relation $d v = d t + d r/\sqrt{F(r) G(r)}$,
which can be solved to obtain an explicit expression for $v$. The metric then takes the form
\begin{eqnarray}
 d s^2 = - G(r) d v^2 + 2 \sqrt{\frac{G(r)}{F(r)}} \, d r  dv + H(r) d \Omega ~.
\end{eqnarray}
Now we allow the mass $m$ in the static solution to depend on the advanced time, $m \to m(v)$. Thereby, we
will assume the mass is zero before an initial value $v_a$ and that the mass stops increasing at $v_b$. We can then, as
before, use the Einstein equations $G=8\pi \widetilde{T}$ to obtain the
effective quantum gravitational stress-energy tensor $\widetilde{T}$. $\widetilde{T}^v_{\; v}$ and $\widetilde{T}^r_{\; r}$
do not change when $m(v)$ is no longer constant. The transverse pressure $\widetilde{T}^\theta_{\; \theta} = \widetilde{T}^\phi_{\;\phi}$
however has an additional term
\beqn
\widetilde{T}^\theta_{\;\theta}(m(v)) = \widetilde{T}^\theta_{\; \theta}(m) -  \frac{P r^2 m^{\prime}(v)}{2 \pi (r + 2 m(v) P)^4}~ ,
\label{transverse}
\eeqn
where $m'=dm/dv$. Because of the ingoing radiation, the stress-energy-tensor now also has an additional non-zero component, $\widetilde{T}^r_{\; v}$, which
describes radially ingoing energy flux
\begin{equation}
G^r_{\; v} = \frac{2 (1 + P)^2 r^4 (r^4 - a_0^2) (r - r_*(v) ) m^{\prime}(v)}{(a_0^2 + r^4)^2 (r + r_*(v))^3}~ .
\end{equation}
Notice that also in the dynamical case, trapping horizons still occur
where $g^{rr}=F(r,v)$ vanishes \cite{bhd,bhd2}, so we can continue to use the notation from
the static case just that $r_\pm(v)$ and $r_*(v)$ are now functions of $v$. The $r$-dependence of this component is depicted in Fig \ref{radiation}. 

This metric reduces to the Vaidya solutions at large radius, or for $\epsilon \to 0, a_0 \to 0$.
However, in the usual Vaidya solutions, the ingoing
radiation creates a central singularity. But as we see here, with the quantum gravitational 
correction, the center remains regular.

We note that the ingoing energy flux has two zeros, one at $r=r_*(v)$ and one
at $r=\sqrt{a_0}$, and is negative between these.
What happens
is that the quantum gravitational correction works against the ingoing flux
by making a negative contribution until the effective flux has dropped to zero
at whatever is larger, the horizon's geometric mean $r_*$ or the location of the dual radius $r=\sqrt{a_0}$.
The flux then remains dominated by the quantum gravitational
effects, avoiding a collapse, until it has passed $r_*$ and the dual radius where it quickly approaches what looks
like an outgoing energy flux to the observer in the second asymptotic region.

Since
in the second asymptotic region $A,A'$ the mass assigned to the white hole
is inversely proportional to the ADM mass at $r=\infty$, the white hole's mass
must be decreasing, consistent with the outgoing (or rather
throughfalling) energy flux. In this process, the past horizon will move
towards smaller $R$ or larger $r$, respectively.

\section{Thermodynamics}
\label{thermo}

Let us now briefly summarize the findings about the thermodynamical
properties of this black hole solution, discussed in more detail in \cite{Modesto:2009ve}.

Particle creation can take place at the horizons $r_+$ and $r_-$ where there
is high blueshift when tracing back light\-rays. However, if the
vacuum at ${\cal I}^-$ in the black hole's asymptotic region $B,B'$ is empty of
particles as usual, then there will be no flux from particle creation at $r_-$ to ${\cal I}^+$ in
the second asymptotic region $A,A'$. This is a consequence of causality and energy
conservation, which we can see as follows.

Consider
there was a particle creation at $r_-$ resulting in a flux of Hawking
radiation towards $R=\infty$. The background is the time-reversed black
hole situation but the flux is not time-reversed. This would mean a decrease of
the white hole's mass for the observer at $R=\infty$. However, since
our metric is geodesically complete, the particles emitted at the white
hole's horizon $r_-$ can be traced back all the way to ${\cal I}^-$ in the black 
hole's asymptotic region $B,B'$. We recall that the white hole's mass
for the observer in the $A,A'$ region is inversely proportional to
the black hole's mass and see that this particle creation at $r_-$ would
contribute to an increase of the black hole's mass 
corresponding to the decrease of the white hole's mass. Since there is
particle emission also at the other horizon $r_+$, we would have
to add both fluxes to obtain the net mass change.

However, we do as usual have a choice for the initial vacuum state
at ${\cal I^-}$ and we will assume as normally that the vacuum in
the black hole's asymptotic past is empty. From the above explanation
we see now that this can only be the case if there is no particle
flux from $r_-$ to the white hole's asymptotic region ${\cal I}^+$. To achieve
this, we have to chose the vacuum at ${\cal I}^-$ in the white
hole's asymptotic region $A,A'$ such that it contains a constant flux
into the white hole with the effect that there is no outgoing particle flux created
at $r_-$. This is the time-reversed situation of an evaporating 
black hole with an empty ingoing vacuum. This situation is
mathematically consistent because particle production in the curved 
background only tells us the relation between the ingoing and outgoing
vacuum states, but not the vacuum states themselves. We thus chose the
vacuum state at ${\cal I}^-$ in the white hole's asymptotic region $A,A'$ 
such that at $r_-$ there is no additional outgoing flux created \footnote{Alternatively,
we could demand the vacuum at past infinity in the second asymptotic region to be
free of particles, but then the vacuum in the black hole region's past infinity would have
to contain particles. We will not further consider this possibility here.}.

Thus, the evaporation proceeds through the Hawking emission at $r_+$, and
the black hole's Bekenstein-Hawking temperature, given in terms of the surface gravity
$\kappa$ by $T_{BH}= \kappa/2 \pi$, yields \cite{Modesto:2009ve}
\begin{eqnarray}
T_{BH}(m) = \frac{(2m)^3 (1 - P^2)}{4 \pi [ (2 m)^4 + a_0^2]} ~.
\label{Temperatura}
\end{eqnarray}
This temperature coincides with the Hawking temperature in the limit of
 large masses but goes to zero for $m \rightarrow 0$.

The luminosity can be estimated by use of the
Stefan-Boltzmann law  $L(m)= \alpha A_H(m) T_{BH}^4(m)$,
where (for a single massless field with two degrees of freedom)
$\alpha = \pi^2/60$, and $A_H(m) = 4 \pi [ (2m)^2+ a_0^2/(2 m)^2]$ is the surface
area of the horizon. Inserting
the temperature, we obtain
\begin{eqnarray}
L(m) =
\frac{16 \, m^{10} \alpha \,(1-P^2)^4}{\pi^3
(a_0^2+16\, m^4)^3}~ .
\label{lumini}
\end{eqnarray}
The mass loss of the black hole is given by $-L(m)$,
\begin{eqnarray}
\frac{d m(v)}{d v} = - L[m(v)]
\label{flux}
\end{eqnarray}
and we can integrate its inverse to obtain
the mass function $m(v)$. The result of this integration with initial
condition $m(v = 0) = m_0$ is
\begin{eqnarray}
&& \hspace{-1cm} v  =  \frac{5 a_0^6 + 432 a_0^4 m^4 + 34560 a_0^2 m^8 - 
   61440 m^{12}) \pi^3}{720 m^9 (1 - P^2)^4 \alpha} \nonumber \\
 && \hspace{-0.7cm} -  \frac{5 a_0^6 + 432 a_0^4 m_0^4 + 34560 a_0^2 m_0^8 - 
   61440 m_0^{12}) \pi^3}{720 m_0^9 (1 - P^2)^4 \alpha}~.
\label{v(m)}
\end{eqnarray}
In the limit $m \rightarrow 0$ this expression becomes $ v \approx a_0^6 \pi^3/(144 m^9 (1 - P^2)^4 \alpha)$,
and one thus concludes that the black hole needs an infinite amount of
time to completely evaporate.

\section{Collapse and Evaporation}
\label{evap}

We are now well prepared to combine formation and evaporation of the
black hole. As in section \ref{dynamical}, we divide space-time into
regions of advanced time. We start with empty space before $v_a$, let
the mass increase from $v_a$ to $v_b$, and stop the increase thereafter.
Hawking radiation will set in, but for astrophysical black holes
this evaporation will proceed very slowly, such that we have a long time span during
which the black hole is quasi-stable and $m$ remains constant to good
accuracy at $m_0$. Then, at some later time, $v_c$, Hawking radiation becomes
relevant and $m$ decreases until it reaches zero again. As we have seen in
the previous section, it will reach zero only in the limit $v\to \infty$.

We thus have the partition $-\infty<v_a<v_b< v_c<\infty$ with
\begin{eqnarray}
\forall v\in(-\infty,v_a)&:&m(v)=0,\label{va}\\
\forall v\in(v_a,v_b)&:&d/dv~ m(v)>0, \label{vb}\\
\forall v\in(v_b,v_c)&:&m(v)=m_0,  \label{vc}\\
\forall v\in(v_c, + \infty)&:&d/dv~ m(v)<0, \label{vd}\\
{\rm for} \,\, v  \rightarrow + \infty
&:&m(v) \rightarrow 0.\label{ve}
\end{eqnarray}
Strictly speaking the mass would immediately start dropping without incoming
energy flux and thus $v_a=v_b$, but stretching this region out will be more illuminating to
clearly depict the long time during which the hole is quasistable.

To describe the Hawking-radiation we will consider the creation of
(massless) particles on the horizon such that locally energy is conserved.
We then have an ingoing radiation with negative energy
balanced by outgoing radiation of positive energy. Both fluxes originate
at the horizon and have the same mass profile which is given by the Hawking
temperature. The area with ingoing negative density is again described
by an ingoing Vaidya solution, while the one with outgoing positive
density is described by an outgoing Vaidya solution.

The outgoing Vaidya solution has a mass-profile that depends on
the retarded time $u$ instead of $v$ and the mass decreases instead
of increases. The retarded time is defined by $du = d t - d r/\sqrt{F(r) G(r)}$. After
a coordinate transformation, the metric reads
\begin{equation}
ds^2= -G(r,u) du^2 - 2 \sqrt{\frac{G(r,u)}{F(r,u)}} du dr + H(r) d\Omega~,
\end{equation}
where $F(r,u)$ and $G(r,u)$ have the same form as in the static case (\ref{statgmunu}),
 but with where $m$ is replaced by a function $m(u)$.
We fix the zero point of the retarded time $u$ so that $r=r_+$
corresponds to $u_c=v_c$. Then there is a static region with total
mass $m_0$ for $v>v_c$, $u<u_c$. Note that since the spacetime described here
has neither a singularity nor an event horizon, we can
consider pair creation to happen directly at the trapping horizon instead
of at a different timelike hypersurface outside the horizon,
as done in \cite{His}. We have in this way further partitioned spacetime in regions,
broken down by retarded time:
\begin{eqnarray}
\forall u<u_c&:&m(u)=m_0 ~, \\
\forall u>u_c&:&d/du~ m(u) <0 ~.
\end{eqnarray}

\begin{figure}
\includegraphics[width=8cm]{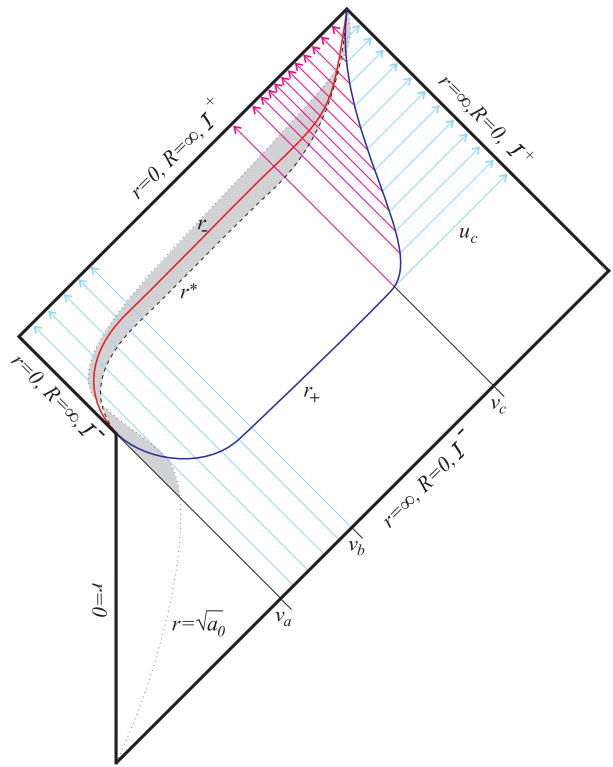}
\caption{Penrose diagram for the formation and evaporation of the regular black hole
metric. The red and dark blue solid lines depict the two trapping horizons $r_-$ and $r_+$. The
brown, dotted line is the curve of $r=\sqrt{a_0}$ and the black, short dashed one is $r_*$.
The light blue arrows represent positive energy flux, the magenta arrows negative energy flux.} \label{bbh}
\end{figure}

Now that we have all parts together, let us explain the complete dynamics
as depicted in the resulting causal diagram Fig.\ref{bbh}.

In the region $v<v_a$ we have a flat and empty region, described
by a piece of Minkowski-space. For all times $v > v_a$, the inner and outer trapping
horizons are present. These horizons join
smoothly at $r=0$ in an infinite time and enclose a non-compact region of trapped surfaces. 

A black hole begins to form at $v=v_a$ from null dust which has collapsed completely at $v=v_b$ to a
static state with mass $m_0$. It begins to evaporate at $v=v_c$, and the complete
evaporation takes an infinite amount of time. The observer at ${\cal I}^+$ sees particle
emission set in at some retarded time $u_c$ which corresponds to the lightlike
surface where the horizon has lingered for a long time. The region with $v>v_c$ is then
divided into a static region for $u<u_c$, and the dynamic Vaidya region for $u>u_c$, which
is further subdivided into an ingoing and an outgoing part.

As previously mentioned, the radially ingoing flux (light blue arrows) in the collapse region is
not positive everywhere due to the quantum gravitational contribution. It has
a flipped sign in the area between $r_*$ (black short dashed curve) and $r=\sqrt{a_0}$ (brown dotted
curve) which is grey shaded in the figure. Likewise, the ingoing negative flux during
evaporation (magenta arrows) has another such region with flipped sign. It is in this region,
between the two horizon's geometric mean value $r_*$ and the dual radius corresponding to the minimal area, that the quantum
gravitational corrections noticeably modify the classical and semi-classical
case, first by preventing
the formation of a singularity, and then by decreasing the black hole's temperature
towards zero.

\section{Conclusions}

We have investigated a model for collapse and evaporation of a black hole that
is entirely singularity-free. The spacetime does not have an event
horizon, but two trapping horizons. By generalizing the previously derived
static metric to a dynamical one by use of the Vaidya metric we found that the
gravitational stress-energy tensor builds up a negative contribution that violates
the positive energy condition and prevents the formation of a singularity. We
divided spacetime into six different regions described by different metrics,
and constructed the causal diagram for the complete evaporation. The value of
the scenario studied here is that it provides a concrete, calculable, model
for how quantum gravitational effects alter the black hole spacetime.

\section*{Acknowledgements}

LM is extremely grateful for the fantastic environment offered by Perimeter Institute.
Research at Perimeter Institute is supported by the Government of Canada through Industry Canada
and by the Province of Ontario through the Ministry of Research \& Innovation.

\section*{APPENDIX} 

The effective energy momentum tensor is defined by ${\widetilde T}^{\mu}_{\; \nu} = G^{\mu}_{ \; \nu}/8 \pi$.
The components of the Einstein tensor in coordinate $(v, r, \theta, \varphi)$ are:
\begin{eqnarray}
&& \hspace{-0.4cm} G^v_{\; v} = \frac{r^2}{(a_0^2 + r^4)^3 (r + 
   r_*)^3} \times \nonumber \\
   && \hspace{-0.4cm} [2 a_0^2 r^4 (r + r_*) (6 r^2 + 7 r_- r_+ - 7 r (r_-  +  r_+) - 2 r r_* \nonumber \\
 && \hspace{-0.4cm}    - r_*^2) 
       - a_0^4 (-r_-  r_+  r_* + r_*^3  + 2 r^2 (r_- + r_+ + 2 r_*) \nonumber \\
       && \hspace{-0.4cm} 
    +  3 r (-r_- r_+ + r_*^2)) 
-  r^8 (-r_- r_+ r_* + r_*^3  \nonumber \\
&& \hspace{-0.4cm} + r (r_- r_+ + 2 (r_- + r_+) r_* + 3 r_*^2))] ~, \nonumber \\
&&\hspace{-0.4cm} G^r_{\; r} = - \frac{r^2}{(a_0^2 + r^4)^3 (r + r_*)^3} \times \nonumber \\
&& \hspace{-0.4cm}   [2 a_0^2 r^4 (r + r_*) (-r_- r_+ + r_*^2 + r (r_- + r_+ + 2 r_*)) \nonumber \\
&& \hspace{-0.4cm}   + a_0^4 (4 r^3 - 2 r^2 (r_- + r_+  - 2 r_*) - r_- r_+ r_* + r_*^3 \nonumber \\
&& \hspace{-0.4cm}    +       r (r_- r_+ + 3 r_*^2)) + 
    r^8 (4 r^2 r_* + 3 r_- r_+ r_* + r_*^3 \nonumber \\
&& \hspace{-0.4cm}    +    r (r_- r_+ - 2 (r_- + r_+) r_* + 3 r_*^2))]~, \nonumber \\
&&\hspace{-0.4cm} G^{\theta}_{\; \theta} = \frac{r^3}{(a_0^2 + r^4)^3 (r + r_*)^4} \times \nonumber \\
 && \hspace{-0.4cm}   [ r^7 (r^2 r_- r_+   + 
       r (2 r^2 + 6 r_- r_+ - 3 r (r_- + r_+)) r_*  \nonumber \\
 && \hspace{-0.4cm}      + (r - 2 r_-) (r - 
          2 r_+) r_*^2) + 2 a_0^2 r^4 (r^2 (r_- + r_+)  \nonumber \\
 && \hspace{-0.4cm}      +    (r_- + r_+) r_*^2 +
       r (r_- - r_*) (-r_+ + r_*))  \nonumber \\
&& \hspace{-0.4cm}  +  a_0^4 (4 r^3 - 2 r^2 (r_-  + r_+   -   3 r_*) + 2 r_- r_+ r_*  \nonumber \\
&& \hspace{-0.4cm} + r (r_- r_+ - 3 (r_- + r_+) r_* + r_*^2))]~.
\end{eqnarray}

\end{document}